\documentclass[twocolumn,prl,showpacs]{revtex4}
\usepackage{graphicx}
\usepackage{color}
\usepackage{dcolumn}

\usepackage{tabularx}

\usepackage{float}

\usepackage{amsmath}
\newcommand{\comment}[1]{}

\begin{document}

%\preprint{}

\title{Low temperature line-width broadening in optical-conductivity spectra of the off-center rattling phonons in type-I clathrate Ba$_8$Ga$_{16}$Sn$_{30}$} 
\author{T. Mori, K. Iwamoto, S. Kushibiki, H. Honda, H. Matsumoto and N. Toyota} 
%\email{mori-t@ldp.phys.tohoku.ac.jp} 
%\homepage{http://ldp.phys.tohoku.ac.jp}
\affiliation{Department of Physics, Graduate School of Science, Tohoku University, Sendai 980-8578, Japan}
\author{K. Suekuni$^1$, M. A. Avila$^1$, and T. Takabatake$^{1,2}$}
\affiliation{$^1$Department of Quantum Matter, ADSM, $^2$Institute for Advanced Material Research, Hiroshima University, Higashi-Hiroshima 739-8530, Japan}

\date{\today}
\begin{abstract}
With a terahertz time-domain spectrometer (0.3 - 3.0 THz) we have measured the optical conductivity of the type-I clathrate Ba$_8$Ga$_{16}$Sn$_{30}$ at temperatures from 300 K down to 7 K. Independent six spectra superimposed on the Drude conductivity are identified to infrared active vibrational modes of guest Ba ions and the cages. While the spectra of five higher-frequency modes depend hardly on temperature, the lowest-lying spectrum with a peak at 0.72 THz due to the Ba(2) ion's off-centering vibration in the oversized cage changes with temperature characteristically. With lowering temperature, the spectral shape of this so-called rattling phonon continues to become so broad that the line-width amounts to be comparable to the peak frequency. Furthermore, below about 100 K, the single broad peak tends to split into two subpeaks. While this splitting can be explained by assuming a multi-well anharmonic potential, the strong enhancement of the line-width broadening toward low temperature, cannot be understood, since the Boltzmann factor generally sharpens the low-temperature spectra. 
\end{abstract}

%\pacs{74.20.Mn,74.25.Gz,74.72.-h}

\maketitle
% body of paper here

%\section{Introduction}

Recently novel phenomena related with local anharmonic phonons observed in cage-like materials such as clathrates\cite{Avila2008,Suekuni2008} have attracted much interest. In particular, in relation to their potential applications to thermoelectric devices, it has been studied extensively how the so-called \textit{rattling phonons} coined for the local anharmonic phonons of guest ions in polyhedral cage could be responsible for such electric and thermal conductivities and the specific heat.

In order to clarify the dynamical properties of these phonons, we reported in the previous paper\cite{Mori2009} the optical-conductivity spectra in the terahertz frequency region for the type-I clathrate Ba$_8$Ga$_{16}$Ge$_{30}$ (BGG), i.e., a $p$-type semiconductor with a low carrier-density. It has turned out that the vibrations of Ba$^{2+}$(2) ions in the tetrakaidecahedra framework are quasi on-centering in an anharmonic potential, indicating characteristic features as the softening of the peak frequency and the sharpening of the line-width toward low temperature\cite{Matsumoto2009,Mori2009}.
These two features of the on-center rattlings are explained by a local anharmonic potential (AP) model. An on-center type anharmonic potential has a non-equivalent level spacing contrary to the harmonic potential; the level spacing becomes larger as the level becomes higher. As the effect of the Boltzmann factor, many and higher levels participate to optical transitions at higher temperature, so that the peak frequency becomes higher and the total line-width becomes broader.
On the other hand, at low temperature, local phonons are almost condensed at the ground level and hence the relevant optical excitation energy becomes lower accompanied with a suppression of the line-width\cite{Matsumoto2009}.

As an extreme case of anharmonicity in the local potential, one can consider an off-centered anharmonic potential. In general such a potential has multiple local minima and may cause the low-lying rattling vibrations with a large displacement. Through couplings with those modes, one expects strong effects on electrons and acoustic phonons. In fact, the type-I clathrate BGS, which is isostructural\cite{Kasper1965, Avila2008} to BGG, is found to show the off-centered displacement of Ba$^{2+}$(2) by the single-crystal x-ray diffraction measurement\cite{Avila2008}. Moreover there have been revealed anomalous properties such as a glass-like behavior in the specific heat and thermal conductivity\cite{Avila2008,Suekuni2008} as well, and a resistivity minimum\cite{Suekuni2008}.
Those behaviors have been discussed in relation to the off-centered rattling phonons studied by Raman scattering measurement\cite{Suekuni2010}. Infrared-active optical conductivity measurements, complementary to the Raman scattering, are expected to shed more light into the dynamical behaviors. Theoretically, in Ref. \cite{Matsumoto2009} for example, it is predicted that the off-centered anharmonic potential would produce a line splitting in the optical conductivity at low temperature together with a softening and/or hardening of the peak frequency.

In this letter, we measure the optical conductivity of infrared-active optical modes of type-I BGS by the terahertz time-domain spectroscopy (THz-TDS) to directly identify the off-centering rattling phonon and reveal its dynamical properties. We discuss the observed results based on the AP model\cite{Matsumoto2009}. The observed spectra, which show the line-splitting just as expected from the model, reveal an unexpected, drastic line-width broadening to occur towards low temperature, which is quite opposite from the usual thermal behavior.

%\section{Experimental}

Single crystals of the type-I BGS doped $n$-type carriers are grown by a self-flux method\cite{Avila2008, Suekuni2008}. To obtain sufficient transmitting signals through the conductive sample, single crystal disks of 3\,-\,5\,mm in diameter are polished down to about 13\,$\mu$m in thickness by using diamond abrasive films. The present measurements covering a frequency range of 0.3 - 3.0\,THz are carried out with the commercial spectrometer (RT-20000, Tochigi Nikon Co. Ltd) which uses a standard technique for the transmission configuration and the temperature range from 7\,K to room temperature\cite{Mori2008, Mori2009}. 

%\section{Results and Discussions}

\begin{figure}
\includegraphics[width=8cm]{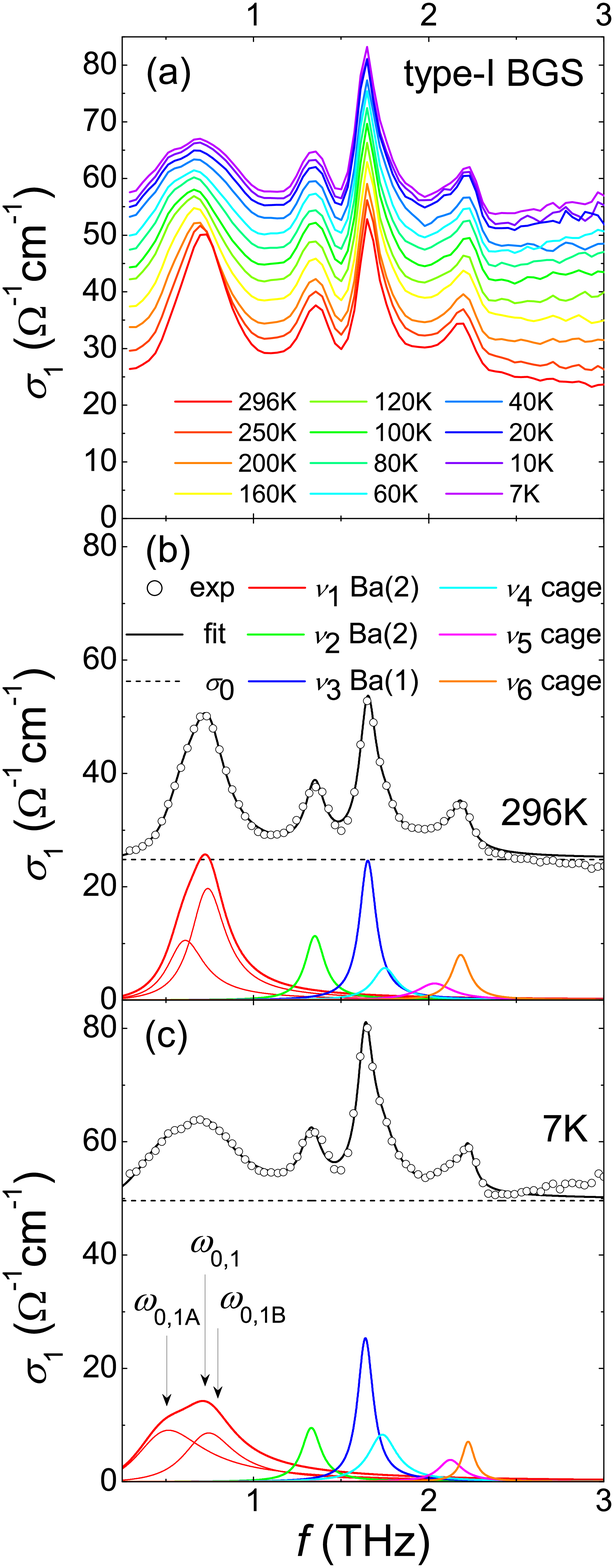}
\caption{(color online) (a) Temperature-dependent real part of complex conductivity spectra $\sigma_{1}(\omega)$ of type-I BGS. (Note the horizontal axis is frequency (= $\omega/2\pi$) in THz). (b) and (c) Experimental data (open circles) and fitting results (lines) at room temperature and 7\,K. Details are described in text. For convenience, 1\,THz = 33.3\,cm$^{-1}$= 48\,K= 4.14\,meV.} 
\label{fig1}
\end{figure}

Figure \ref{fig1} shows the real part, $\sigma_1(\omega)$, of the complex conductivity for data at temperatures from room temperature down to the lowest 7\,K. Just as previously found in the isostructural BGG\cite{Mori2009}, the data demonstrate four broad peaks superimposed on almost frequency-independent Drude-like contributions (see horizontal dotted lines in (b) and (c)) caused by doped carriers. The latter contribution monotonically increases with decreasing temperature. This metallic behavior is contrary to the semiconducting behavior of $p$-type BGG in terahertz frequency region\cite{Mori2009}. 

The observed four peaks are assigned as six independent modes ($\nu_1-\nu_6$) by the first-principle calculations for the infrared active optical modes of $T_{1u}$ symmetry\cite{Suekuni2010}. To reproduce each spectral shape, we define the fitting curve as
\begin{equation}
\sigma_{1,fit}(\omega)=\sigma_{0}+\sum_{i}\frac{(2/\pi)S_{i}\omega^{2}\Gamma_{i}}{(\omega_{0,i}^2-\omega^2)^2+(\omega\Gamma_{i})^2},
\label{fitting}
\end{equation}
where the first term $\sigma_0$ is the dc conductivity ($\omega = 0$) of the carrier contribution, and the second term is a Lorentz-type conductivity for the phonon mode $\nu_{i}$. There, $S_{i}$, $\omega_{0,i}$, and $\Gamma_i$ are a spectral weight, a resonant angular frequency and a relaxation rate (or a damping factor), respectively. We apply this equation to fit all the data and demonstrate the fitting results at room temperature and 7\,K in Fig. \ref{fig1}(b) and (c). Fitting parameters are listed in Table \ref{tab1}.

\begin{table}
\caption{Fitting parameters at 296\,K (7\,K) of type-I BGS. The guest ions in the smaller ($2a$) and oversized ($6d$) cages are defined as Ba(1) and (2), respectively. The definitions of $\omega_{0,1}$ and $\Gamma_{1}$ are described in text.}
\label{tab1}
\begin{center}
{\renewcommand\arraystretch{1.2}
\setlength{\arrayrulewidth}{0.2pt}
\begin{tabularx}{0.48\textwidth}{@{\extracolsep{\fill}}ccccc} \hline\hline
$label$ & $ atom$ & $\omega_{0,i}$ & $\Gamma_{i}$ & $S_{i} \times 10^{-13}$ \\
& & $(\textrm{THz})$ & $(\textrm{THz})$ & $(\Omega^{-1}\textrm{cm}^{-1}s^{-1})$ \\ \hline
$\nu_{1}$ & Ba(2) & 0.72 (0.71) & 0.32 (0.57) & 7.75 (7.27) \\
$\nu_{2}$ & Ba(2) & 1.35 (1.33) & 0.14 (0.13) & 2.40 (2.00) \\
$\nu_{3}$ & Ba(1) & 1.65 (1.64) & 0.19 (0.11) & 3.00 (2.76) \\
$\nu_{4}$ & cage & 1.75 (1.74) & 0.15 (0.21) & 0.82 (1.76) \\
$\nu_{5}$ & cage & 1.98 (2.12) & 0.31 (0.18) & 0.93 (0.74) \\
$\nu_{6}$ & cage & 2.18 (2.23) & 0.16 (0.09) & 1.40 (0.65) \\ \hline\hline
\end{tabularx}}
\end{center}
\end{table}

Firstly, the lowest peak labeled $\nu_1$ is assigned as the mode of the Ba$^{2+}$(2) in the oversized tetrakaidecahedra cage. In order to reproduce the line shapes, this peak is fitted by superposing two Lorentzian curves, $L_{1\textrm{A}}$ and $L_{1\textrm{B}}$. The peak frequency of $\omega_{0,1}$ in Table \ref{tab1} is defined as the peak position of the superposed Lorentzian curves (see, the arrows in Fig. \ref{fig1} (c)). The relaxation rate $\Gamma_1$ of $\nu_1$ is evaluated from the difference between the frequencies of half maximum value for the curves. The $\nu_1$ mode corresponds to the vibration within the plane perpendicular to the fourfold inversion axis, i.e., the off-center rattling phonon mode in the oversized cage.

Secondly, the small peak around 1.3\,THz labeled $\nu_2$ is also assigned as the mode of the Ba$^{2+}$(2) in the oversized tetrakaidecahedra cage. This mode is the vibration along the fourfold inversion axis.

Thirdly, the third peak spectrum around 1.7\,THz is reproduced by superposing two Lorentzian curves, where the larger one is assigned as the mode labeled $\nu_3$ of Ba$^{2+}$(1) in the smaller dodecahedra cage and the smaller one as the mode labeled $\nu_4$ of the cage.

Finally, the peak around 2.2\,THz is also fitted by superposition of two Lorentzian curves, and the both curves are assigned as the cage modes labeled $\nu_5$ and $\nu_6$. It is to note that the spectral weight $S_{i}$ for all the modes are independent on temperature within an experimental error.

\begin{figure}
\includegraphics[width=8cm]{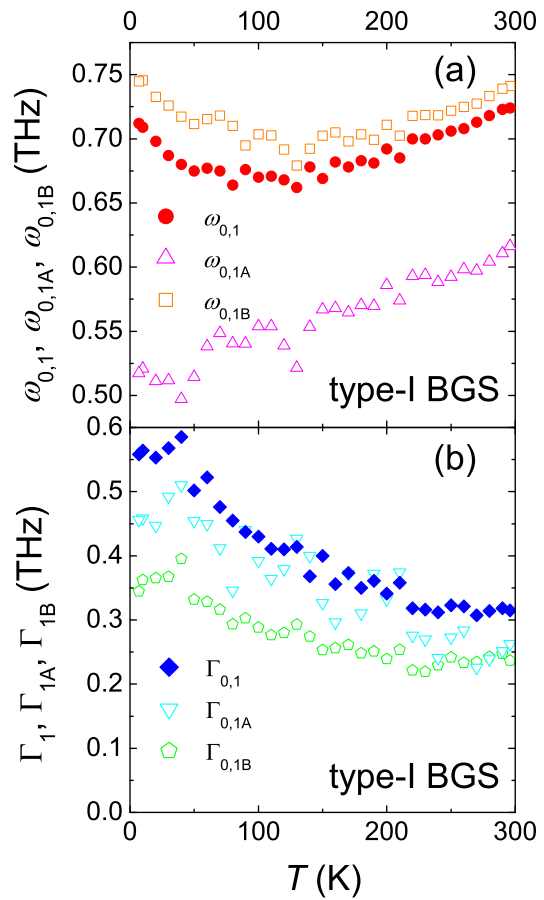}
\caption{(color online) Temperature dependence of (a) the peak frequency $\omega_{0,1}$ (circles) and (b) total line width $\Gamma_{1}$ (diamonds) for $\nu_1$ mode of type-I BGS. The resonant frequencies ($\omega_{0,1\textrm{A}}$ (open triangles) and $\omega_{0,1\textrm{B}}$ (open squares)) and the relaxation rates ($\Gamma_{1\textrm{A}}$ (open inverted triangles) and $\Gamma_{1\textrm{B}}$ (open pentagons)) are evaluated from the fitting for $\nu_1$ mode.}     
\label{fig2}
\end{figure}

As can be seen in Fig. \ref{fig1} (a), all the spectral shapes ($\nu_2-\nu_6$) except the lowest $\nu_1$ do depend little on temperature. For the peak frequencies, the $\nu_2$ mode softens by 1.5\% and the $\nu_3$ mode also softens very slightly as temperature decreases, while other higher modes of cages become constant or hardened. These behaviors are essentially similar to those in BGG\cite{Mori2009}. 

Hereafter we will concentrate on the $\nu_1$ mode of the rattling phonon that is drastically temperature dependent. As seen from Fig. \ref{fig1} (a) and also shown by circles in Fig. \ref{fig2} (a), the peak frequency of $\nu_1$ mode $\omega_{0,1}$ slightly softens by 7\,\% toward 100\,K and turns to harden at further lower temperature. This tendency has never been observed in on-center type rattling phonons in BGG\cite{Mori2009}. The line-width $\Gamma_{1}$ indicated by marked diamonds in Fig. \ref{fig2} (b) increases almost by a factor of 2 from 0.3\,THz at room temperature to 0.6\,THz at 7\,K. 

\begin{figure}
\includegraphics[width=8cm]{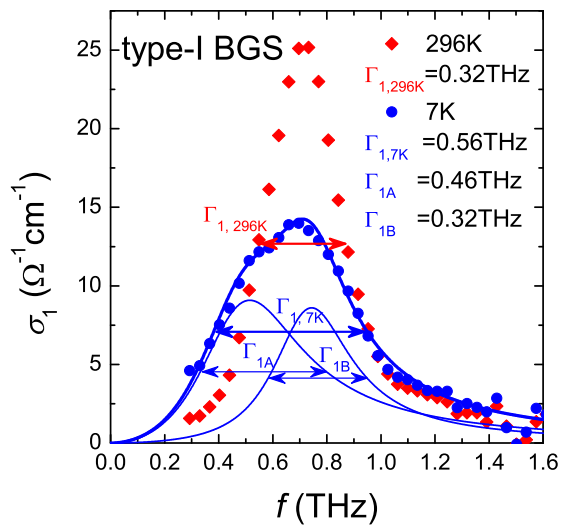}
\caption{(color online) Expanded views for the spectral shapes of $\nu_1$ (in Fig.\ref{fig1}) at room temperature (diamonds) and 7\,K (circles) and fitting curves at 7\,K (lines). Experimental data are analyzed by subtracting the fitting results for neighboring phonon modes ($\nu_2$, $\nu_3$) and the flat Drude conductivity from the raw data. Thin lines are the Lorentzian curves of $L_{1\textrm{A}}$ and $L_{1\textrm{B}}$) with the relaxation rates $\Gamma_{1\textrm{A}}$ and $\Gamma_{1\textrm{B}}$. Thick line with $\Gamma_{1,7\textrm{K}}$ is obtained by superposing these two curves.}
\label{fig3}
\end{figure}

For clarity, Fig. \ref{fig3} expands the spectral shapes at room temperature (diamonds) and 7\,K (circles) already shown in Fig. \ref{fig1}. At 7\, K, the $\nu_1$ spectrum clearly deforms around 0.6\,THz, and has been well reproduced by assuming the double peaks at about 0.5 and 0.7\,THz. 

\begin{figure}
\includegraphics[width=8cm]{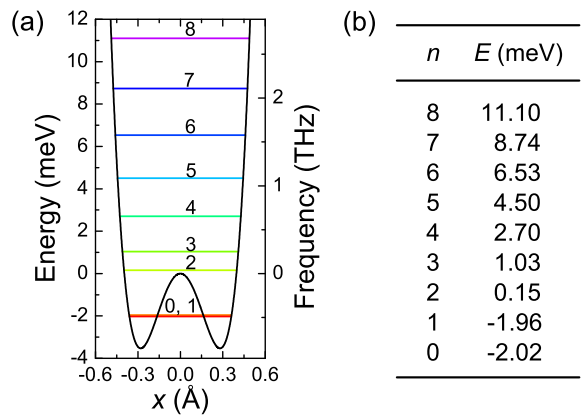}
\caption{(color online) (a) Calculated potential shape $V(x)$ (black line) and eigen values (horizontal bars). $V(x)=1/2kx^2+1/4\lambda x^4$ where $k = -2.5[\rm kgs^{-2}]$ and $\lambda = 2.7 \times 10^{21}[\rm m^{-2}kgs^{-2}]$. The superscript of each eigen level indicates number of each level. Note that level 0 and 1 are almost degenerated. (b) Calculated eigen energies for $n$ = 0 to 8.}
\label{fig4}
\end{figure}

Based on the analysis using 1D-AP model, the splitting at low temperature is well understood using the concrete vibrational levels in the double-well potential $V(x)=1/2kx^2+1/4\lambda x^4$ shown in Fig. \ref{fig4}. The parameters are fixed to reproduce the frequencies, $\omega_{0,1\textrm{A}}$=0.52\,THz and $\omega_{0,1\textrm{B}}$=0.74\,THz for the double peaks. Using the value of the Ba mass, potential parameters are uniquely determined; $k = - 2.9 (\rm kgs^{-2})$ and $\lambda = 3.7 \times 10^{21} (\rm m^{-2}kgs^{-2})$. The features of the calculated levels are as follows. Firstly, the $n = 0$ and 1 levels of are almost degenerated with a tiny energy difference $\hbar\omega_{10}$ = 0.059\,meV. Secondly, the $n = 2$ and 3 levels are well separated due to the moderately deep potential well. From the ground level $n = 0$, $\hbar\omega_{20}=2.17$\,meV (=0.52\,THz) and $\hbar\omega_{30}=3.05$\,meV (=0.74\,THz), respectively. These characteristic low-lying levels from $n = 0$ to 3 can produce double peaks at low temperature as follows. At sufficient low temperatures, phonons are condensed to almost degenerate levels of $n = 0$ and 1. Therefore the optical transitions occur mainly from $n = 0$ to 3 and from $n = 1$ to 2 due to a parity requirement of the wave function. These transitions result in forming the double peaks with the excitation energies, $\omega_{21}\approx\omega_{20}=0.52$\,THz and $\omega_{30}=0.74$\,THz. With these excitations and the Lorentzian curve (eq. 1), the data are best fitted as shown in Fig.\ref{fig3}, leading to $\Gamma_{1\textrm{A}}=0.46$\,THz and $\Gamma_{1\textrm{B}}=0.32$\,THz just, respectively corresponding to $\Gamma_{30}$ and $\Gamma_{21}$.

However, the significant broadening observed in the rattling phonon spectra toward low temperature cannot be reproduced by our calculations based on the 1D-AP model or its extension to more realistic potentials of higher order and dimension. (The details will be published elsewhere.)

In conclusion, the temperature dependence of the conductivity spectra has been determined for the off-centering rattling phonon in the oversized cage of type-I BGS. Using the 1D-AP model, the spectral splitting at low temperatures can be explained, qualitatively at least, as optical excitations from highly degenerate ground and the first excited levels to the second and third ones. Then the spectra are well reproduced by superposing two Lorentzian shapes. There remains a question, however, on the present finding that the line-width, a direct measure for the relaxation rates of the rattling mode, increases continually with decreasing temperature, which would be quite opposite to the spectral sharpening usually expected from the Boltzmann factor.

So far we have discussed our spectra with the 1D-AP and its extended models treating only the anharmonic optical modes. All the other possible interactions are neglected at all; the dipole-dipole interaction between neighboring rattling phonons\cite{Nakayama2008, Kaneshita2009} and the interactions with acoustic phonons propagating along the cage network. The interaction with the conducting charge carriers has also been discussed\cite{Takechi2009}. Since the present spectroscopic technique provides us a phonon-mode selective window, further experimental and theoretical studies are highly encouraged to delineate how the off-center rattlers are interacting with these excitations. 

%\begin{acknowledgements}

We thank T. Hasegawa, Y. Takasu, M. Udagawa, E. Kaneshita, T. Nakayama, K. Ueda, A. Yamakage, Y. Kuramoto, K. Iwasa, S. Iwai and H. Matsui for valuable discussions. Two of us (T. M., K. I.) are supported financially by the Global COE programm ``Materials Integrations'', Tohoku University, and one of us (H. M.) partially by CREST (JST). The works at Sendai have been supported by Grants-in-Aid for Scientific Research (A)(15201019) and the priority area ``Nanospace'' from MEXT, Japan. The works at Higashi Hiroshima have been supported by Grants in Aid for Scientific Research (A)(18204032), the priority area ``Nanospace''(1951011) and ``Skutterudite''(15072205), and the innovative areas ``Heavy Electrons''(20102004) from MEXT, Japan, and the Sasakawa Scientific Research from Japan Science Society. 

%\end{acknowledgements}

\end{document}